\documentclass[]{article}
\usepackage{geometry}
\usepackage{amsmath}
\usepackage{color}
\usepackage{graphicx}
\usepackage{amssymb}
\usepackage{mathrsfs}
\usepackage{bm}
\def\scri{\mathscr{I}}

\title{Dynamic and Thermodynamic Stability of Charged Perfect Fluid Stars}
\author{
\footnotesize Kai Shi$^1$\thanks{Email:kaishi@mail.bnu.edu.cn},
Yu Tian$^2$\thanks{Email: ytian@ucas.ac.cn},
Xiaoning Wu$^3$\thanks{Email: wuxn@amss.ac.cn},
Hongbao Zhang$^1$\thanks{Email: hongbaozhang@bnu.edu.cn},
Jingchao Zhang $^1$\thanks{Email: jczhang@mail.bnu.edu.cn}
\\
\footnotesize $^1$Department of Physics, Beijing Normal University, Beijing 100875, China\\
\footnotesize $^2$ School of Physics, University of Chinese Academy Sciences, Beijing 100049, China\\
\footnotesize $^3$ Institute of Mathematics, Chinese Academy of Sciences, Beijing 100190, China}
\begin{document}
\maketitle

\begin{abstract}
We perform a thorough analysis of the dynamic and thermodynamic stability for the charged perfect fluid star by applying the Wald formalism to the Lagrangian formulation of Einstein-Maxwell-charged fluid system. As a result, we find that neither the presence of the additional electromagnetic field nor the Lorentz force experienced by the charged fluid makes any obstruction to the key steps towards the previous results obtained for the neutral perfect fluid star. Therefore, the criterion for the dynamic stability of our charged star in dynamic equilibrium within the symplectic complement of the trivial perturbaions with the ADM $3$-momentum unchanged is given by the non-negativity of the canonical energy associated with the timelike Killing field, where it is further shown for both non-axisymmetric and axisymmetric perturbations that the dynamic stability against these restricted perturbations also implies the dynamic stability against more generic perturbations. On the other hand, the necessary condition for the thermodynamic stability of our charged star in thermodynamic equilibrium is given by the positivity of the canonical energy of all the linear on-shell perturbations with the ADM angular momentum unchanged in the comoving frame, which is equivalent to the positivity of the canonical energy associated with the timelike Killing field when restricted onto the axisymmetric perturbations. As a by-product, we further establish the equivalence of the dynamic and thermodynamic stability with respect to the spherically symmetric perturbations of the static, spherically symmetric isentropic charged star.

\end{abstract}
\section{Introduction}
Partly motivated by the Gubser-Mitra conjecture on the relationship between the dynamic instability and thermodynamic instability of the black branes in AdS spacetimes\cite{GM,GM2}, Hollands and Wald established a criterion for the dynamic stability of black holes as well as its close connection with the thermodynamic stability in \cite{HW} by working with the Wald formalism\cite{LW,Wald,IW}. Such a strategy was further extended successfully to the relativistic neutral perfect fluid stars\cite{GSW}, where not only were the dynamic and thermodynamic stability  analysed in a transparent and unified manner, demonstrating its very advantage over the case by case explicit calculations\cite{SWZ,Roupas,Fang2017,Fang2018}, but also some new observations were made for the axisymmetric perturbations, compared to the previous explorations\cite{C1970,FS1975,FS1978,F1978,LH}.

Although most of the real life stars are believed to be nearly electrically neutral\footnote{Among other possible exceptions, quark stars are suspected to be generically charged\cite{GL}.}, it is at least of academic interest to inquire about whether the above results for the neutral perfect fluid stars also hold for the charged perfect fluid stars. Moreover, it is noteworthy that such a charged perfect fluid star in an AdS spacetime also plays an important role in addressing the holographic metallic criticality\cite{HT}. But nevertheless, there is no a priori answer to the above inquiry, since the electromagnetic field as well as the Lorentz force acting on the charged perfect fluid kicks in the game. The purpose of this paper is intended to provide a definite answer by performing a comprehensive and detailed analysis of the dynamic and thermodynamic stability of the charged perfect fluid star in an asymptotically flat spacetime.
As a result, we find that neither the presence of the electromagnetic field nor the electromagnetic force experienced by the charged fluid makes any obstruction to the key steps towards the results obtained in \cite{GSW} for the neutral star. To this end, we are especially required to derive an explicit expression of the canonical energy at null infinity for the electromagnetic part and argue for its non-negativity as well as its zero value for the physically stationary perturbations. Among others, we are also required to show that the angular momentum per particle, vorticity, and circulation share the same properties as those for the neutral perfect fluid stars, although their expressions get modified in the presence of the electromagnetic field. Thus we are able to obtain exactly the same results for the dynamic and thermodynamic stability of our charged stars as those presented in \cite{GSW} for the neutral stars. In passing, not only do we provide more detailed derivations and arguments in some places where they are only sketched or completely skipped in \cite{GSW}, but also we present a thorough alternative derivation for Eq. (\ref{zero}) to that suggested in \cite{GSW}.

The rest of this paper is structured as follows. In the next section, we shall provide a brief review of the dynamics of the charged perfect fluid, the definition of the resulting star in dynamic equilibrium, and the criterion for its thermodynamic equilibrium as well as the corresponding first law of thermodynamics. In Section \ref{main}, we devote ourselves to developing the Lagrangian formulation of the Einstein-Maxwell-charged fluid system. To this end, we first set up the dynamical fields for our system, pin down the associated redundancies, define the Eulerian and Lagrangian variations and point out their relations in the subsection \ref{subdynamical}. Then we propose the Lagrangian for our system and apply the Wald formalism to it in the subsection \ref{sublagrangian}, where we particularly introduce the symplectic form and define the three Noether current related charges for both the diffeomorphisms and $U(1)$ gauge transformations. In the subsection \ref{subphase}, we figure out the phase space of our system, resort to the canonical variables to introduce an inner product on the subspace of perturbations, on which the symplectic complement is defined and the double symplectic complement of a subspace is shown to be itself. Finally, we introduce the canonical energy for the linear on-shell perturbations in the subsection \ref{subcanonical}, where not only do we present the explicit expression of the canonical energy at null infinity for both the electromagnetic and gravitational parts, but also show the relation of the canonical energy with the second order perturbations. With such a preparation, we take advantage of the physically stationary perturbations to establish the criterion for the dynamic stability of our charged star within some subspace of linear on-shell perturbations in Section \ref{DS}, where the implication of such a dynamic stability for both non-axisymmetric and axisymmetric perturbations are further explored, respectively. In Section \ref{TS}, we further establish the criterion for the thermodynamic stability of our charged star against generic perturbations as well as axisymmetric perturbations, and show the equivalence of the dynamic and thermodynamic stability with respect to the spherically symmetric perturbations of the static, spherically symmetric isentropic charged star. We conclude our paper with some discussions.

The notations and conventions of \cite{GRbook} will be followed by us, except that the indices are not required to be balanced in our equations if no confusion arises. In particular, early Latin letters $(a,b,c,...)$ denote abstract spacetime indices while middle Latin letters $(i,j,k,...)$ denote concrete spatial indices on a spacelike Cauchy surface unless specified otherwise. In addition, the differential forms with the indices omitted are indicated in bold typeface and a vector contracting with the first index of a differential form is denoted by the dot with it.

\section{Charged perfect fluid star, dynamic equilibrium, and thermodynamic equilibrium}
By a charged perfect fluid in gravitational and electromagnetic fields, we mean that its energy momentum and charge current satisfy the following equations of motion\cite{HT}
\begin{equation}\label{conservationlaw}
    \nabla_aT_m^{ab}=F^{bc}J_c, \quad \nabla_a J^a=0
\end{equation}
with $\bm{F}=d\bm{A}$ the electromagnetic field strength and
\begin{equation}
    T_m^{ab}=(\rho+p)u^au^b+pg^{ab},\quad J^a=nu^a.
\end{equation}
Here the charge of per particle has been set to be unity such that the charge density and number density are conflated, and $u^a$ is the $4$-velocity satisfying $u_au^a=-1$. In addition, the energy density is determined by the equation of state $\rho(n,s)$ and the pressure can be obtained as
\begin{equation}
p=n\frac{\partial\rho}{\partial n}-\rho,
\end{equation}
where we have used Euler's equation
\begin{equation}
    \rho+p=(Ts+\mu)n
\end{equation}
with $s$ the entropy per particle and $\mu$ the chemical potential, as well as the local version of the first law of thermodynamics
\begin{equation}
    d\rho=(Ts+\mu)dn+Tnds.
\end{equation}
Furthermore, one can show that the entropy current is also conserved, i.e., $\nabla_a (sJ^a)=0$, which is equivalent to the statement that the entropy per particle along the $4$-velocity is constant.

The resulting charged star is said to be in dynamic equilibrium if
\begin{equation}
    \mathscr{L}_tA_a=\mathscr{L}_\varphi A_a=\mathscr{L}_tu^a=\mathscr{L}_\varphi u^a=\mathscr{L}_t n=\mathscr{L}_\varphi n=\mathscr{L}_t s=\mathscr{L}_\varphi s=0
\end{equation}
for the timelike and axial Killing vector fields $t^a$ and $\varphi^a$, as well as the $4$-velocity $u^a$ taking the following circular form
\begin{equation}
    u^a=(t^a+\Omega \varphi^a)/|v|
\end{equation}
with $\Omega$ the angular velocity and
\begin{equation}
    |v|^2=-g_{ab}(t^a+\Omega\varphi^a)(t^b+\Omega\varphi^b).
\end{equation}
Note that $v^a\nabla_a(|v|^2)=2v^av^b\nabla_av_b=0$ with $v^a=t^a+\Omega \varphi^a$, we have $\nabla_au^a=0$. Namely, for the charged star in dynamic equilibrium, the corresponding $4$-velocity is divergence free, which further implies that the particle number density along the $4$-velocity is also constant.

As proven in \cite{our}, the above charged star in dynamic equilibrium is in thermodynamic equilibrium if and only if
\begin{equation}
    \tilde{T}\equiv |v|T=\text{const.},\quad \tilde{\mu}\equiv |v|\mu=\text{const.},\quad \Omega=\text{const.}
\end{equation}
throughout the whole star\footnote{Note that we have $\nabla_au^a=0$, $\sigma^{ab}\equiv q^{ac}q^{bd}(\nabla_{(c}u_{d)}-\frac{1}{3}g_{cd}\nabla_eu^e)=0$, and $\nabla_a(\frac{\mu}{T})=0$ for a star in thermodynamic equilibrium,  so its energy momentum tensor, charge current, and entropy current receive no correction individually from the bulk viscosity, shear viscosity, and electric conductivity in the first order gradient expansion of dissipative hydrodynamics as it should be the case\cite{Kovtun}.}. Moreover, the charged star in thermodynamic equilibrium satisfies the first law of thermodynamics as
\begin{equation}
    \delta \mathcal{M}=\tilde{T}\delta S+\tilde{\mu}\delta N+\Omega \delta \mathcal{J}
\end{equation}
with $\mathcal{M}$ the ADM mass,  $\mathcal{J}$ the ADM angular momentum, $S$ the total entropy, and $N$ the total particle number.

\section{Lagrangian formulation of the Einstein-Maxwell-charged fluid system}\label{main}
\subsection{Dynamical fields, redundancies, and variations}\label{subdynamical}
To develop a Lagrangian description of the Einstein-Maxwel-charged fluid system, not only are we required to have the spacetime manifold $M$, on which the metric $g_{ab}$ and the electromagnetic potential $A_a$ are defined, but also we like to introduce a fiducial manifold $M'$, called fluid spacetime, which is diffeomorphic to $M$. Then with a fixed scalar field $s'$ and a fixed $3$-form $\bm{\mathcal{N}}'$ on $M'$, satisfying
\begin{equation}
    d\bm{\mathcal{N}}'=d(s'\bm{\mathcal{N}}')=0,
\end{equation}
one can define the physical fluid fields on $M$ by pushing forward with a diffeomorphism $\chi$ as
\begin{equation}
	nu\cdot\bm{\epsilon}\equiv\bm{\mathcal{N}}=\chi_*\bm{\mathcal{N}}',\quad 	s=\chi_*s',
\end{equation}\
where $\bm{\epsilon}$ is the associated spacetime volume element. Accordingly, both the charge current conservation and the entropy current conservation are automatically satisfied.

With the above prescription, we can take $\phi=(g_{ab},A_a,\chi)$ as
the dynamical fields for our Einstein-Maxwell-charged fluid system, where $\chi$ can be understood equivalently as a collection of 4 scalar fields $x'^{\mu}\circ\chi^{-1}$ on $M$ with $x'$ the local coordinates on $M'$. But nevertheless, it is noteworthy that such a description has an additional redundancy besides the usual diffeomorphism and $U(1)$ gauge redundancies. Namely, two field configurations $\phi=(g_{ab},A_a,\chi)$ and $\tilde{\phi}=(g_{ab},A_a,\tilde{\chi})$ are physically equivalent to each other, or trivially related if $\chi_*\bm{\mathcal{N}}'=\tilde{\chi}_*\bm{\mathcal{N}}'$ and $\chi_*s'=\tilde{\chi}_*s'$, which amounts to saying that $\bm{\mathcal{N'}}$ and $s'$ are invariant under $\tilde{\chi}^{-1}\circ\chi$, or equivalently $\bm{\mathcal{N}}$ and $s$ are invariant under $\tilde{\chi}\circ\chi^{-1}$.

The variation around an arbitrary field configuration $\phi$ can be formulated by introducing a one-parameter family of dynamical fields $\phi(\lambda)=(g_{ab}(\lambda),A_a(\lambda),\chi_\lambda)$ with $\phi(0)=\phi$. Note that $\varphi_\lambda\equiv\chi_\lambda\circ\chi_0^{-1}$ gives rise to a one-parameter family of diffeomorphisms on $M$, so the first order perturbation is completely specified by a triple
\begin{equation}
    \delta\phi
\equiv(\delta g_{ab}\equiv\frac{d g_{ab}(\lambda)}{d\lambda}|_{\lambda=0},\delta A_a\equiv\frac{dA_a(\lambda)}{d\lambda}|_{\lambda=0},\xi^a\equiv\frac{dx^\mu\circ\varphi_\lambda}{d\lambda}(\frac{\partial}{\partial x^\mu})^a|_{\lambda=0}),
\end{equation}
where the vector field $\xi^a$ is known as the Lagrangian displacement with $x$ the local coordinate on $M$. A first order perturbation $\delta\mathcal{Q}$ of an arbitrary tensor field $\mathcal{Q}$ induced by $\delta\phi$ is usually called the Eulerian perturbation. More generally, the $k^\text{th}$-order Eulerian perturbation of $\mathcal{Q}$ is defined as
\begin{equation}
    \delta^k\mathcal{Q}\equiv\frac{d^k\mathcal{Q}(\lambda)}{d\lambda^k}|_{\lambda=0}.
\end{equation}
However, it turns out to be convenient to consider the so called Lagrangian perturbation, in which the variation is performed onto the diffeomorphism equivalent field configuration $\hat{\phi}(\lambda)=(\varphi_\lambda^*g_{ab}(\lambda),\varphi_\lambda^*A_a(\lambda),\chi_0)$. Put it another way, the $k^\text{th}$-order Lagrangian perturbation of $\mathcal{Q}$ is defined as
\begin{equation}
    \Delta^k\mathcal{Q}\equiv \frac{d^k\hat{\mathcal{Q}}(\lambda)}{d\lambda^k}|_{\lambda=0}
\end{equation}
with $\hat{\mathcal{Q}}(\lambda)=\varphi_\lambda^*\mathcal{Q}(\lambda)$.
Whence we obviously have
\begin{equation}
    \Delta^k\phi=(\Delta^kg_{ab},\Delta^kA_a,0), \quad \Delta^k\bm{\mathcal{N}}=0, \quad \Delta^k s=0
\end{equation}
for any $k\geq 1$. In addition, it is easy to show
\begin{equation}
    \Delta\mathcal{Q}=\delta\mathcal{Q}+\mathscr{L}_\xi\mathcal{Q}
\end{equation}
for the first order perturbation, whereby we further have
\begin{equation}
\delta \bm{\mathcal{N}}=-\mathscr{L}_\xi\bm{\mathcal{N}}, \quad \delta s=-\mathscr{L}_\xi s.
    \end{equation}
$\xi^a$ is called a trivial displacement if $\delta\bm{\mathcal{N}}=0$ and $\delta s=0$. In what follows, we shall denote such a trivial displacement as $\eta^a$. It is not hard to show that $\eta^a$ can always be written in the form of the superposition of two trivial displacements as follows\cite{GSW}
\begin{equation}
    \eta^a=fu^a+\tilde{\eta}^a,
\end{equation}
where $fu^a$ is called the flowline trivial with $f$ an arbitrary function and
\begin{equation}
    \tilde{\eta}=\frac{1}{n^2}\mathcal{N}^{abc}\nabla_bZ_c,
\end{equation}
with $Z_c$ satisfying
\begin{equation}
    u^cZ_c=0,\quad \mathscr{L}_uZ_c=0,\quad \nabla_{[a}s\nabla_bZ_{c]}=0.
\end{equation}
Furthermore, one can show $\bm{Z}$ can be expressed as
\begin{equation}\label{special}
    \bm{Z}=Fds
\end{equation}
with $\mathscr{L}_u F=0$ for $\nabla_as\neq0$ and as a sum of terms of the form $F_1dF_2$
with $\mathscr{L}_uF_1=\mathscr{L}_uF_2=0$ for $\nabla_as=0$.
\subsection{Lagrangian, symplectic form, and Noether current related charges}\label{sublagrangian}
Next we like to apply the Wald formalism to the Einstein-Maxwell-charged fluid system with the Lagrangian $4$-form given by
\begin{equation}
	\bm{L}=R\bm{\epsilon}-\frac{1}{4}F_{ab}F^{ab}\bm{\epsilon}+A_aJ^a\bm{\epsilon}-\rho\bm{\epsilon}
\end{equation}
with $J^a=nu^a$.
Note that
\begin{equation}
\begin{aligned}
    	\delta (R\bm{\epsilon})&=-G^{ab}\delta g_{ab}\bm{\epsilon}+ d\bm{\theta}_{g},\\
    	\delta (-\frac{1}{4}F_{ab}F^{ab}\bm{\epsilon})&=\nabla_bF^{ba}\delta A_a\bm{\epsilon}+\frac{1}{2}T^{ab}_{EM}\delta g_{ab}\bm{\epsilon}+d\bm{\theta}_{EM}
    	\end{aligned}
\end{equation}
with the energy momentum tensor for the electromagnetic field given by
\begin{equation}
    T_{EM}^{ab}=F^{a}{}_cF^{bc}-\frac{1}{4}F^{cd}F_{cd}g^{ab}
\end{equation}
and
\begin{equation}
\begin{aligned}
\bm{\theta}_g&=g^{ab}g^{cd}(\nabla_d\delta g_{bc}-\nabla_b\delta g_{cd})\epsilon_{aefg}, \\
\bm{\theta}_{EM}&=-F^{ab}\delta A_b\epsilon_{aefg}.
\end{aligned}
\end{equation}
In addition, we can derive
\begin{equation}
	\begin{aligned}
	     \delta(A_aJ^a\bm{\epsilon})&=(\Delta-\mathscr{L}_\xi)(J^aA_a\bm{\epsilon})\\
		&=\Delta A_aJ^a\bm{\epsilon}-\mathscr{L}_\xi(A_aJ^a\bm{\epsilon})\\
		&=J^a\delta A_a\bm{\epsilon}+J^a\mathscr{L}_\xi A_a\bm{\epsilon}-d(J^aA_a\xi\cdot\bm{\epsilon})\\
		&=J^a\delta A_a\bm{\epsilon}+J^a\xi^bF_{ba}\bm{\epsilon}+\nabla_a(\xi^bA_bJ^a)\bm{\epsilon}-d(A_aJ^a\xi\cdot\bm{\epsilon})\\
		&=J^a\delta A_a \bm{\epsilon}+J^a\xi^b F_{ba}\bm{\epsilon}+d\bm{\theta}_{m1}
	\end{aligned}
\end{equation}
with
\begin{equation}
\bm{\theta}_{m1}=\xi^aA_aJ\cdot\bm{\epsilon}-A_aJ^a\xi\cdot\bm{\epsilon},
\end{equation}
where we have used the Cartan identity $\mathscr{L}_w\bm{F}=w\cdot d\bm{F}+d(w\cdot\bm{F})$ as well as the fact that
\begin{equation}
	\begin{aligned}
		\Delta \bm{\epsilon}&=\frac{1}{2}g^{ab}\Delta g_{ab}\bm{\epsilon}\\
		\Delta u^a&=\frac{1}{2}u^au^bu^c\Delta g_{bc}\\
		\Delta n&=-\frac{1}{2}nq^{bc}\Delta g_{bc}.
	\end{aligned}
\end{equation}
in the second step with $q^{bc}=g^{bc}+u^bu^c$ and $\nabla_aJ^a=0$ in the fourth step. With the help of $\Delta\rho=\frac{\rho+p}{n}\Delta n$ and by the same token, one can show
\begin{equation}
\begin{aligned}
    \delta (-\rho\bm{\epsilon})&=\frac{1}{2}T^{ab}_m\delta g_{ab}\bm{\epsilon}-\xi_b\nabla_aT_m^{ab}\bm{\epsilon}+d\bm{\theta}_{m2}
\end{aligned}
\end{equation}
with the energy momentum tensor for our fluid taking the familiar form
\begin{equation}
    T_m^{ab}=(\rho+p)u^au^b+pg^{ab}
\end{equation}
and
\begin{equation}
    \bm{\theta}_{m2}=(\xi_bT_m^{ba}+\rho\xi^a)\epsilon_{aefg}.
\end{equation}
Therefore the variation of the full Lagrangian $4$-form can be encapsulated as
\begin{equation}
    \delta\bm{L}=\bm{E}(\phi)\delta\phi+d\bm{\theta}(\phi;\delta\phi).
\end{equation}
Here $\bm{E}=0$ gives rise to the following equations of motion
\begin{equation}
	 \begin{aligned}
E^{ab}&\equiv	 -G^{ab}+\frac{1}{2}T^{ab}=0,\\
E_{EM}^a&\equiv	 	\nabla_bF^{ba}+J^a=0,\\
E_{mb}&\equiv	 -	\nabla^aT_{mab}+F_{ba}J^a=0
	 \end{aligned}
\end{equation}
with the total energy momentum tensor $T^{ab}=T_{EM}^{ab}+T_m^{ab}$, and the total pre-symplectic potential $3$-form is given by
\begin{equation}
    \bm{\theta}=\bm{\theta}_g+\bm{\theta}_{EM}+\bm{\theta}_{m}
\end{equation}
with
\begin{equation}
	\bm{\theta}_{m}=\bm{\theta}_{m1}+\bm{\theta}_{m2}=\xi^aP_{abcd},
\end{equation}
where we have defined
\begin{equation}
    P_{abcd}=[(\rho+p)q_a{}^e-A_fJ^f \delta_a{}^e +A_fJ^e\delta_a{}^f]\epsilon_{ebcd}=[(\rho+p)q_a{}^e-A_fJ^f q_a{}^e +A_fJ^eq_a{}^f]\epsilon_{ebcd}.
\end{equation}
Now with $\delta\phi$ formally viewed as a vector at the point $\phi$ in the configuration space $\mathcal{F}$, denoted as $\delta\phi^A$, one can define
a $1$-form field $\Theta_A$ and the pre-symplectic form on $\mathcal{F}$ as follows
\begin{equation}
    \Theta_A\delta\phi^A=\int_\Sigma\bm{\theta}(\phi;\delta\phi),\quad
    \Omega_{AB}=(D\Theta)_{AB},
\end{equation}
where $\Sigma$ denotes a Cauchy surface in $M$ and $D$ represents the exterior derivative on $\mathcal{F}$. A convenient way to evaluate  $\Omega_{AB}\delta_1\phi^A\delta_2\phi^B$ at the point $\phi\in\mathcal{F}$ is to extend $\delta_1\phi^A$ and $\delta_2\phi^A$ off of $\phi$ in an arbitrary manner and use the following formula
\begin{equation}
    \Omega_{AB}\delta_1\phi^A\delta_2\phi^B=\mathscr{L}_{\delta_1\phi}(\Theta_A\delta_2\phi^A)-\mathcal{L}_{\delta_2\phi}(\Theta_A\delta_1\phi^A)-\Theta_A[\delta_1\phi,\delta_2\phi]^A.
\end{equation}
This amounts to saying
\begin{equation}
      \Omega_{AB}\delta_1\phi^A\delta_2\phi^B=\int_\Sigma \bm{\omega}(\phi;\delta_1\phi,\delta_2\phi),
\end{equation}
with the pre-symplectic current $3$-form on $M$ defined as
\begin{equation}
\bm{\omega}(\phi;\delta_1\phi,\delta_2\phi)=\delta_1\bm{\theta}(\phi;\delta_2\phi)-\delta_2\bm{\theta}(\phi;\delta_1\phi)-\bm{\theta}(\phi;\delta_1\delta_2\phi-\delta_2\delta_1\phi).
\end{equation}
For our purpose, we choose $(\delta_1g_{ab},\delta_1A_a)$ and $(\delta_2g_{ab},\delta_2A_a)$ as variations along a two-parameter family of metrics and gauge fields $(g_{ab}(\lambda_1,\lambda_2),A_a(\lambda_1,\lambda_2))$, which implies
\begin{equation}
    \delta_1\delta_2g_{ab}=\delta_2\delta_1g_{ab},\quad \delta_1\delta_2A_a=\delta_2\delta_1A_a.
\end{equation}
On the other hand, we like to keep $\xi_1^a$ and $\xi_2^a$ fixed, i.e.,
\begin{equation}
    \delta_1\xi_2^a(x)=\delta_2\xi_1^a(x)=0,
\end{equation}
which implies
\begin{equation}
    \delta_1\delta_2\chi=\delta_1\xi_2^a(\chi)=\xi_1^b\partial_b\xi_2^a(\chi),\quad     \delta_2\delta_1\chi=\delta_2\xi_1^a(\chi)=\xi_2^b\partial_b\xi_1^a(\chi).
\end{equation}
Whence we have
\begin{equation}
    \delta_1\delta_2\chi-\delta_2\delta_1\chi=[\xi_1,\xi_2]^a.
\end{equation}
As a result, the pre-symplectic form can be written explicitly as
\begin{eqnarray}\label{sf}
    \Omega_{AB}\delta_1\phi^A\delta_2\phi^B&=&\int_\Sigma (\delta_2h_{ij}\delta_1\bm{\pi}^{ij}-\delta_1h_{ij}\delta_2\bm{\pi}^{ij})+\int_\Sigma(\delta_2A_i\delta_1\bm{\pi}^i-\delta_1A_i\delta_2\bm{\pi}^i)\nonumber\\
    &&+\int_\Sigma(\xi_2^a\delta_1P_{aefg}-\xi_1^a\delta_2P_{aefg}-[\xi_1,\xi_2]^aP_{aefg}),
\end{eqnarray}
where
\begin{equation}
   \bm{\pi}^{ij}=(K^{ij}-Kh^{ij})\hat{\bm{\epsilon}}, \quad \bm{\pi}^i=\nu_aF^{ai}\hat{\bm{\epsilon}}
\end{equation}
with $K_{ij}$ the extrinsic curvature  and $\hat{\bm{\epsilon}}=\nu\cdot\bm{\epsilon}$ the induced volume on $\Sigma$ by the future directed normalized normal vector $\nu^a$.

But nevertheless, no matter whether $\delta_1\delta_2\phi-\delta_2\delta_1\phi$ vanishes or not, one can show that we always have
\begin{equation}
    d\bm{\omega}(\phi;\delta_1\phi,\delta_2\phi)=\delta_2\bm{E}\delta_1\phi-\delta_1\bm{E}\delta_2\phi,
\end{equation}
which means that $\bm{\omega}$ is closed if both $\delta_1\phi$ and $\delta_2\phi$ satisfy the linearized equations of motion $\delta_1\bm{E}=\delta_2\bm{E}=0$. Accordingly, $\Omega_{AB}\delta_1\phi^A\delta_2\phi^B$ does not depends on the choice of the Cauchy surface $\Sigma$ if not only do $\delta_1\phi$ and $\delta_2\phi$ satisfy the linearized equations of motion but also have an appropriate fall-off behavior at the spatial infinity. In addition, for the local symmetry associated with the diffeomorphisms generated by an arbitrary vector field $X^a$, we can define the corresponding Noether current $3$-form as
\begin{equation}
    \bm{J}_X=\bm{\theta}(\phi;\mathscr{L}_X\phi)-X\cdot\bm{L},
\end{equation}
with $\mathscr{L}_X\phi=(\mathscr{L}_Xg_{ab},\mathscr{L}_XA_a,-X^a)$, whereby we have
\begin{equation}
    d\bm{J}_X=-\bm{E}\mathscr{L}_X\phi.
\end{equation}
Therefore the corresponding Noether charge
\begin{equation}
    Q_X=\int_\Sigma \bm{J}_X=0
\end{equation}
 for the on-shell field configurations provided that $X^a$ vanishes sufficiently rapidly at the spatial infinity. Furthermore, one can show
 \begin{equation}
     \bm{\omega}(\phi;\delta\phi,\mathscr{L}_X\phi)=\delta\bm{J}_X+X\cdot\bm{E}(\phi)\delta\phi-d(X\cdot\bm{\theta}(\phi,\delta\phi)).
 \end{equation}
 Note that the Noether current can be written as follows
\begin{equation}
    \bm{J}_X=X^a\bm{C}_a+d\bm{Q}_X
\end{equation}
for our Einstein-Maxwell-charged fluid system with
\begin{equation}
   \bm{ C}_a=(-2E^b{}_a-E_{EM}^bA_a)\epsilon_{befg},\quad \bm{Q}_X=-*d\bm{X}-*\bm{F}A_cX^c,
\end{equation}
where the star denotes the Hodge dual. So we end up with the following fundamental identity
\begin{equation}\label{fi}
 \bm{\omega}(\phi;\delta\phi,\mathscr{L}_X\phi)=X^b\delta\bm{C}_b+X\cdot\bm{E}(\phi)\delta\phi+d(\delta \bm{Q}_X-X\cdot\bm{\theta}(\phi,\delta\phi)).
\end{equation}
One immediate implication of this identity is the diffeomorphism invariance of the symplectic form. Speaking specifically, if $\phi$ is on-shell, $\delta\phi$ satisfies the linearized constraint equations $\delta\bm{C}_a=0$, and $X^a$ vanishes sufficiently rapidly as before, we have
\begin{equation}
    \Omega_{AB}\delta\phi^A\mathscr{L}_X\phi^B=0,
\end{equation}
which amounts to saying that $\Omega_{AB}\delta_1\phi^A\delta_2\phi^B$ keeps invariant under the gauge shift $\delta\phi\rightarrow \delta\phi+\mathscr{L}_X\phi$. On the other hand, for a large gauge transformation, where $X^a$ approaches a nontrivial asymptotic symmetry, we instead have
\begin{equation}\label{si}
    \Omega_{AB}\delta\phi^A\mathscr{L}_X\phi^B=\int_\Sigma(X\cdot\bm{E}\delta\phi+ X^a\delta\bm{C}_a)+\int_{S_\infty}(\delta\bm{Q}_X-X\cdot\delta \bm{B}),
\end{equation}
 where we have assumed that
\begin{equation}
    \int_{S_\infty}X\cdot\bm{\theta}(\phi;\delta\phi)=\int_{S_\infty}X\cdot\delta\bm{B}(\phi)
\end{equation}
for some $3$-form $\bm{B}$. By evaluating Eq. (\ref{si}) at an on-shell $\phi$, we have
\begin{equation}
    \Omega_{AB}\delta\phi^A\mathscr{L}_X\phi^B=\delta H_X=\delta\phi^AD_AH_X
\end{equation}
with the Hamiltonian conjugate to $X^a$ defined as
\begin{equation}
    H_X=\int_\Sigma X^a\bm{C}_a+\int_{S_\infty}(\bm{Q}_X-X\cdot\bm{B}).
\end{equation}
By further restricting ourselves onto the phase space $\mathcal{P}$, which is obtained by factoring out the degeneracy orbits of $\Omega_{AB}$ in the configuration space $\mathcal{F}$, we wind up with a non-degenerate $\Omega_{AB}$. Thus we have the familiar form of Hamilton's equation on the phase space as
\begin{equation}
    \mathscr{L}_X\phi^A=\Omega^{AB}D_BH_X
\end{equation}
with $\Omega^{AB}$ the inverse of $\Omega_{AB}$, satisfying $\Omega^{AB}\Omega_{BC}=\delta^A{}_C$.

The ADM charge $\mathcal{Q}_X$ is defined as the surface term in $H_X$, which is equal to $H_X$ when evaluated at an on-shell $\phi$. In particular, for the asymptotic time translation $t^a$ and rotation $\varphi^a$, the corresponding ADM mass and angular momentum are defined as
\begin{equation}
\mathcal{M}=\int_{S_\infty}(\bm{Q}_t-t\cdot\bm{B})=\int_{S_\infty}(-*d\bm{t}-t\cdot \bm{B}),\quad \mathcal{J}=-\int_{S_\infty}\bm{Q}_\varphi=\int_{S_\infty}*d\bm{\varphi},
\end{equation}
where not only have we chosen $S_\infty$ such that $\varphi^a$ is tangent to it but also have taken the gauge such that $A_a|_{S_\infty}=0$. If the rotation $\varphi^a$ is a background symmetry generator, the ADM angular momentum can be expressed as follows
\begin{equation}
\begin{aligned}
    \mathcal{J}&=-2\int_\Sigma \nabla_a\nabla^{[a}\varphi^{b]}\epsilon_
{befg}=-2\int_\Sigma \nabla_a\nabla^a\varphi^b\epsilon_{befg}=2\int_\Sigma R^{ab}\varphi_a\epsilon_{befg}=\int_\Sigma T^{ab}\varphi_a\epsilon_{befg}\\
&=\int_\Sigma [\frac{\rho+p}{n}\varphi\cdot\bm{u}\bm{\mathcal{N}}+F^{cb}\nabla_c(\varphi\cdot\bm{A})\epsilon_{befg}]=\int_\Sigma [\varphi\cdot(\frac{\rho+p}{n}\bm{u}+\bm{A})\bm{\mathcal{N}}-d*\bm{F}(\varphi\cdot\bm{A})]\\
&=\int_\Sigma \varphi\cdot(\frac{\rho+p}{n}\bm{u}+\bm{A})\bm{\mathcal{N}}=\int_\Sigma j\bm{\mathcal{N}}
\end{aligned}
\end{equation}
with $\Sigma$ so chosen that $\varphi^a$ is tangent to it and $j\equiv\varphi\cdot(\frac{\rho+p}{n}\bm{u}+\bm{A})$ interpreted as the angular momentum per particle. Here we have used $d*\bm{\Lambda}=-\nabla_a\Lambda^{ab}\epsilon_{befg}$ for any $2$-form $\bm{\Lambda}$ in the first step, $\nabla_a\nabla_b\kappa_c=R_{cba}{}^d\kappa_d$ for any Killing vector field $\kappa^a$ in the third step, and $\mathscr{L}_\varphi\bm{A}=0$ as well as the Cartan identity in the fifth step.

Similarly, associated with the local symmetry under the field configuration independent $U(1)$ gauge transformation $\delta_\vartheta\phi=(0,\nabla_a\vartheta,0)$, we can define the corresponding Noether current as
\begin{equation}
    \bm{J}_\vartheta=\bm{\theta}(\phi;\delta_\vartheta\phi)-\vartheta\bm{\mathcal{N}},
\end{equation}
whereby we have
\begin{equation}
    \begin{aligned}
        \delta\bm{J}_\vartheta&=\bm{\omega}(\phi;\delta\phi,\delta_\vartheta\phi)+\xi^a\delta_\vartheta\bm{P}_a-\vartheta\delta\bm{\mathcal{N}}\\
        &=\bm{\omega}(\phi;\delta\phi,\delta_\vartheta\phi)+\xi^a[-\nabla_f(\vartheta J^f)\epsilon_{abcd}+\nabla_a\vartheta\bm{\mathcal{N}}]+\vartheta\mathscr{L}_\xi\bm{\mathcal{N}}\\
        &=\bm{\omega}(\phi;\delta\phi,\delta_\vartheta\phi)-\xi\cdot d(\vartheta\bm{\mathcal{N}})+\mathscr{L}_\xi(\vartheta\bm{\mathcal{N}})\\
        &=\bm{\omega}(\phi;\delta\phi,\delta_\vartheta\phi)+d(\vartheta\xi\cdot\bm{\mathcal{N}})
    \end{aligned}
\end{equation}
In addition, a straightforward calculation yields
\begin{equation}
    \bm{J}_\vartheta=\vartheta\bm{C}+d\bm{Q}_\vartheta
\end{equation}
with
\begin{equation}
    \bm{C}=-E^{EM}\cdot\bm{\epsilon},\quad \bm{Q}_\vartheta=-*\bm{F}\vartheta.
\end{equation}
Thus we obtain
\begin{equation}
    \bm{\omega}(\phi;\delta\phi,\delta_\vartheta\phi)=\vartheta\delta\bm{C}+d(\delta\bm{Q}_\vartheta-\vartheta\xi\cdot\bm{\mathcal{N}}),
\end{equation}
whereby one can obviously introduce the analogous charges and make the analogous statements to those following Eq. (\ref{fi}). In particular, note that $d*\bm{F}=\bm{\mathcal{N}}$, so we have the on-shell variation $\delta H_\vartheta=0$ automatically for those gauge transformations which preserve the gauge condition $A_a=0$ at $S_\infty$.

\subsection{Phase space, inner product, and symplectic complement}\label{subphase}
As alluded to in the previous subsection, the phase space is obtained by factoring out the degeneracy orbits in the configuration space. Put it another way,  $\delta\phi_\text{p}=0$ is the only degeneracy of the symplectic form if and only if one can parameterize the phase space by $\phi_\text{p}$. To proceed, we would like first to introduce the space of fiducial flowlines $\Sigma'$, namely the space of the integral curves of a non-vanishing $u'^a$ with $u'\cdot\bm{\mathcal{N}}'$. Then we can further introduce the diffeomorphism $\psi$ from $\Sigma'$ to $\Sigma$ obtained by intersecting with $\Sigma$ the images of the fiducial flowlines under $\chi$. Next we shall show that one can parameterize the phase space for our Einstein-Maxwell-charged fluid system by $\phi_\text{p}=(h_{ij},\bm{\pi}^{ij},A_i,\bm{\pi}^i,\psi, u^i)$ on $\Sigma$ with  $u^i=h^{ia}u_a$ the fluid $3$-velocity, namely we are required to show that
$ \Omega_{AB}\delta_1\phi^A\delta_2\phi^B=0$ for any $\delta_2\phi$ necessitates $\delta_1\phi_\text{p}=0$. To this end, we like to express
\begin{equation}\label{sfd}
\begin{aligned}
    \Omega_{AB}\delta_1\phi^A\delta_2\phi^B&=\Omega_{AB}\delta_1\phi^A\Delta_2\phi^B-\Omega_{AB}\delta_1\phi^A\mathscr{L}_{\xi_2}\phi^B\\
    &=\int_\Sigma[(\Delta_2h_{ij}\delta_1\bm{\pi}^{ij}-\delta_1h_{ij}\Delta_2\bm{\pi}^{ij}+\Delta_2A_i\delta_1\bm{\pi}^i-\delta_1A_i\Delta_2\bm{\pi}^i-\xi_1^a\Delta_2\bm{P}_a)- (\xi_2\cdot\bm{E}\delta_1\phi+\xi_2^a\delta_1\bm{C}_a)],
    \end{aligned}
\end{equation}
where we have used Eq. (\ref{sf}) to evaluate the first term and Eq. (\ref{si}) together with the Lagrangian displacement of spatial compact support for the second term.
By $\Delta q_a{}^e=u^eq_a{}^{(b}u^{c)}\Delta g_{bc}$, one can show
\begin{equation}
    	\Delta \bm{P}_a=\bm{A}_a{}^{bc}\Delta g_{bc}+\bm{B}_a{}^f{}\Delta A_f
\end{equation}
with
\begin{equation}
	\bm{A}_a{}^{bc}=-\frac{1}{2}(\rho+p)u^bu^cq_a{}^e\epsilon_{ebcd}-\frac{1}{2}c_s^2(\rho+p)q^{bc}q_a{}^e\epsilon_{ebcd}+\frac{\rho+p}{n}q_a{}^{(b}u^{c)}\bm{\mathcal{N}},\quad
		\bm{B}_{a}{}^{f}=2q_{a}{}^{[f}J^{e]}\epsilon_{ebcd},
\end{equation}
where $c_s$ is the sound speed, defined as $c_s^2\equiv(\frac{\partial p}{\partial\rho})_s=\frac{\Delta p}{\Delta \rho}$ and assumed to satisfy $0\le c^2_s\le1$. Furthermore, with the $3+1$ decomposition in the coordinates $(\tau,x^i)$ with $\Sigma$ given by $\tau=0$, we have
\begin{equation}
    ds^2=-\alpha^2d\tau^2+h_{ij}(dx^i+\beta^id\tau)(dx^j+\beta^jd\tau),\quad \bm{A}=Ad\tau+A_i(dx^i+\beta^id\tau),
\end{equation}
whereby we can express
\begin{equation}
    \Delta g_{ab}=-\frac{2}{\alpha}\nu_a\nu_b\Delta\alpha-\frac{1}{\alpha}\nu_a\Delta N_b-\frac{1}{\alpha}\nu_b\Delta N_a+\Delta\gamma _{ab},\quad \Delta A_a=-\frac{1}{\alpha}\nu_a\Delta\mathcal{A}+\Delta \mathcal{A}_a
\end{equation}
with $\Delta N_a=\Delta N_i(dx^i+\beta^id\tau)_a\equiv h_{ij}\Delta\beta^j(dx^i+\beta^id\tau)_a$, $\Delta \gamma_{ab}\equiv \Delta h_{ij}(dx^i+\beta^id\tau)_a(dx^j+\beta^jd\tau)_b$, $\Delta\mathcal{A}\equiv\Delta A+A_i\Delta\beta^i$, and $\Delta\mathcal{A}_a\equiv\Delta A_i(dx^i+\beta^id\tau)_a$. The degeneracy of our symplectic form is obtained by requiring the coefficients of $\Delta \bm{\pi}^{ij}$, $\Delta_2\alpha$, $\Delta_2 N_i$, $\Delta_2 h_{ij}$, $\Delta_2\bm{\pi}^i$, $\Delta_2\mathcal{A}$, $\Delta_2 A_i$, and $\xi_2^a$ in Eq. (\ref{sfd}) vanish when pulled back onto $\Sigma$, which gives rise to
\begin{equation}\label{degeneracy}
    \begin{aligned}
        \delta_1h_{ij}&=0,\\
        \xi_1^a\bm{A}_a{}^{bc}\nu_b\nu_c&=0,\\
        \xi_1^a\bm{A}_a{}^{bi}\nu_b&=0,\\
        \delta_1\bm{\pi}^{ij}-\xi_1^a\bm{A}_a{}^{ij}&=0,\\
        \delta_1A_i&=0,\\
        \xi_1^a\bm{B}_a{}^b\nu_b&=0,\\
        \delta_1\bm{\pi}^i-\xi_1^a\bm{B}_a{}^i&=0,\\
        \delta_1\bm{C}_a-\nu\cdot\bm{E}\delta_1\phi\nu_a&=0.
    \end{aligned}
\end{equation}
Here the second and third equations can be encapsulated as
\begin{equation}
0=\xi_1^a\bm{A}_a{}^{bc}\nu_b=-\frac{1}{2}(\rho+p)[(u^b\nu_b)^2\delta^c{}_d-c_s^2\nu_bq^{bc}\nu_d]\xi_1^aq_a{}^d\hat{\bm{\epsilon}}.
\end{equation}
The contraction with $\nu_c$ implies $\nu_d\xi_1^aq_a{}^d=0$, whereby one can further obtain $\xi_1^aq_a{}^d=0$ by the above equation. That is to say  $\xi_1^a\propto u^a$, namely $\delta_1\psi=0$. In addition, by $u^a\bm{A}_a{}^{bc}=u^a\bm{B}_a{}^b=0$, the fourth and seventh equations in Eq. (\ref{degeneracy}) yield $\delta_1\bm{\pi}^{ij}=\delta_1\bm{\pi}^i=0$. So right now we are only left with $\delta u^i=0$ to show. To achieve this, we first decompose $\delta_1\phi$ as follows
\begin{equation}
  ( \delta_1g_{ab}, \delta_1A_a, \xi_1^a)=(0, 0, fu^a)+(\delta_1g_{ab}, \delta_1A_a, \tau\zeta^a)
\end{equation}
with $u^a\zeta_a=0$. Note that the last equation in Eq. (\ref{degeneracy}) is automatically satisfied by $(0, 0, fu^a)$ since  $u^aE_{ma}=0$ due to the built-in conservation of the charge current and entropy current in our Lagrangian description. Furthermore, $\delta_1h_{ij}=\delta_1\bm{\pi}^{ij}=\delta_1A_i=\delta_1\bm{\pi}^i=0$ implies that the pure gravitational and electromagnetic parts do not contribute to the left side of the last equation in Eq. (\ref{degeneracy}). Accordingly, for $(\delta_1g_{ab},\delta_1A_a,\tau\zeta^a)$, we have
\begin{equation}\label{lengthy}
\begin{aligned}
0&=-\delta_1(T^b_{ma}\epsilon_{befg})-\frac{1}{2}T_m^{bc}\delta_1g_{bc}\nu_a\hat{\bm{\epsilon}}-\delta_1(A_a\bm{\mathcal{N}})-J^b\delta_1A_b\nu_a\hat{\bm{\epsilon}}\\
&=-\delta_1(\frac{\rho+p}{n}u_a\bm{\mathcal{N}}+p\epsilon_{aefg})-\frac{1}{2}[(\rho+p)u^bu^c\delta_1g_{bc}+pg^{bc}\delta_1g_{bc}]\nu_a\hat{\bm{\epsilon}}-\delta_1(A_a\bm{\mathcal{N}})-J^b\delta_1A_b\nu_a\hat{\bm{\epsilon}}\\
&=-\delta_1(\frac{\rho+p}{n}u_a)\bm{\mathcal{N}}+\delta_1p\nu_a\hat{\bm{\epsilon}}+\frac{1}{2}(\rho+p)u_bu_c\delta_1g^{bc}\nu_a\hat{\bm{\epsilon}}-\delta_1A_a\bm{\mathcal{N}}-J^b\delta_1A_b\nu_a\hat{\bm{\epsilon}}\\
&=[nu_b\nu^b\Delta_1(\frac{\rho+p}{n}u_a)+\Delta_1p\nu_a-(\rho+p)u_b\nu^bu_c\delta_1\nu^c\nu_a]\hat{\bm{\epsilon}}\\
&=-(\rho+p)(\frac{1}{2}c_s^2u_b\nu^bq^{cd}\Delta_1g_{cd}u_a-u_b\nu^b\Delta_1u_a+\frac{1}{2}c_s^2q^{cd}\Delta_1g_{cd}\nu_a+u_b\nu^bu_c\delta_1\nu^c\nu_a)\hat{\bm{\epsilon}}\\
&=-(\rho+p)(\frac{1}{2}c_s^2q^{cd}\Delta_1g_{cd}q_{ab}\nu^b-u_b\nu^b\Delta_1u_a+u_b\nu^bu_c\delta_1\nu^c\nu_a)\hat{\bm{\epsilon}}\\
&=-(\rho+p)[c_s^2(\nu_c\delta_1\nu^c+u_c\nu^cu_d\delta_1\nu^d-\frac{1}{\alpha}\zeta_c\nu^c)q_{ab}\nu^b-u_b\nu^b\Delta_1u_a+u_b\nu^bu_c\delta_1\nu^c\nu_a]\hat{\bm{\epsilon}},
\end{aligned}
\end{equation}
where we have used $\delta_1\bm{\mathcal{N}}=-\mathscr{L}_{\tau\zeta}\bm{\mathcal{N}}=-d\tau\wedge\zeta\cdot\bm{\mathcal{N}}=0$ when restricted onto $\Sigma$ in the third step, $\delta_1h^{ab}=0$ as well as $\delta_1A_a\propto\nu_a$ in the fourth step, and $\Delta_1g_{ab}=\delta_1g_{ab}-\frac{1}{\alpha}(\nu_a\zeta_b+\nu_b\zeta_a)$
in the seventh step. To proceed, we compute
\begin{equation}\label{key1}
\begin{aligned}
    \Delta_1u_a&=\nu_au_d\delta_1\nu^d+u_d\nu^dg_{ac}\delta_1\nu^c-\frac{1}{\alpha}\zeta_au_d\nu^d+u_au_c\nu^cu_d\delta_1\nu^d\\
    &=h_{ab}\delta_1(h^{bc}u_c)+\nu_a[u_d\delta_1\nu^d-u_d\nu^d\nu_c\delta_1\nu^c+\frac{1}{\alpha}\zeta_c\nu^cu_d\nu^d-(u_c\nu^c)^2u_d\delta_1\nu^d],
    \end{aligned}
\end{equation}
whereby we further have
\begin{equation}\label{key2}
    u_b\delta_1(h^{bc}u_c)=u^ah_{ab}\delta_1(h^{bc}u_c)=(u_c\nu^c)^2(\nu_d\delta_1\nu^d-\frac{1}{\alpha}\zeta_d\nu^d+u_b\nu^bu_d\delta_1\nu^d).
\end{equation}
Then plugging Eq. (\ref{key1}) into Eq. (\ref{lengthy}) and using Eq. (\ref{key2}), we end up with
\begin{equation}\label{zero}
0=-(\rho+p)B_{ab}\delta_1(h^{bc}u_c)\hat{\bm{\epsilon}},
\end{equation}
where
\begin{equation}
    B_{ab}=\frac{c_s^2}{(u_d\nu^d)^2}q_{ac}\nu^cu_b-2u^ch_{b[a}\nu_{c]}.
\end{equation}
By contracting Eq. (\ref{zero}) with $\nu^a$, we obtain $u_b\delta_1(h^{bc}u_c)=0$, whereby Eq. (\ref{zero}) further implies $\delta_1(h^{ab}u_b)=0$. Thus we have accomplished the proof that the phase space for our Einstein-Maxwell-charged fluid system is described exactly by $\phi_\text{p}$ on $\Sigma$.

However, the variables $(\psi,u^i)$ are not canonically conjugate. To construct the canonically conjugate variables for our charged fluid, we like to choose the coordinates in $M'$ such that the fiducial flowlines are given by the integral curves of $(\frac{\partial}{\partial x'^0})^a$. Then by $\delta x'^\mu=-\partial_a x'^\mu\xi^a$, we
can write
\begin{equation}
    \bm{\theta}_m=-\delta x'^\mu \bm{P}'_\mu=-\delta x'^i\bm{P}'_i
\end{equation}
with $\bm{P}'_\mu=(\frac{\partial}{\partial x'^\mu})^a\bm{P}_a$. Whence we further have
\begin{equation}
    \bm{\omega}_m=-(\delta_2x'^i\delta_1\bm{P}'_i-\delta_1x'^i\delta_2\bm{P}'_i),
\end{equation}
which tells us that  $(x'^i, -\bm{P}'_i)$ can be regarded as the canonically conjugate variables for our charged fluid. Accordingly, the symplectic form for our Einstein-Maxwell-charged fluid system  can be cast into the following canonical manner
\begin{equation}
    \Omega_{AB}\delta_1\phi^A\delta_2\phi^B=\int_\Sigma(\delta_2q^\alpha\delta_1p_\alpha-\delta_1q^\alpha\delta_2p_\alpha)
\end{equation}
with $q^\alpha=(h_{ij},A_i, x'^i)$ and $p_\alpha=(\bm{\pi}^{ij},\bm{\pi}^i,-\bm{P}'_i)$. Then by working with the coordinates in which $h=1$, we can introduce an inner product
\begin{equation}
    \langle\delta_1\phi,\delta_2\phi\rangle=\int_\Sigma(h_{\alpha\beta}\delta_1q^\alpha\delta_2q^\beta+h^{\alpha\beta}\delta_1p_\alpha\delta_2p_\beta),
\end{equation}
where $h_{\alpha\beta}$ and $h^{\alpha\beta}$ should be understood as either $h_{ij}$ or $h^{ij}$ depending on the tensor indices of $(q^\alpha, p_\alpha)$.
Associated with this inner product, we have the Hilbert space $\mathcal{H}$ as a subspace of perturbations. As pointed out in \cite{GSW}, the square integrability indicates that $\mathcal{H}$ does not include those perturbations with $\delta \mathcal{M}\neq0$, but includes all the perturbations of interest with $\delta\mathcal{M}=0$. Then it is not hard to see that our symplectic form can be expressed as a bounded linear operator $W$ on $\mathcal{H}$ as follows
\begin{equation}
    \Omega_{AB}\delta_1\phi^A\delta_2\phi^B=\langle \delta_1\phi,\hat{\Omega}\delta_2\phi)
\end{equation}
with
\begin{equation}
    \hat{\Omega}(\delta q^\alpha,\delta p_\alpha)=(-h^{\alpha\beta}\delta p_\beta,h_{\alpha\beta}\delta q^\beta).
\end{equation}
Whence we know that $\hat{\Omega}$ is an orthogonal map, since $\hat{\Omega}^2=-1$ and $\hat{\Omega}^\dagger=-\hat{\Omega}$.

The symplectic complement of a subspace $\mathcal{S}$ in $\mathcal{H}$ is defined as follows
\begin{equation}
    \mathcal{S}^{\perp_\text{s}}=\{\delta'\phi\in \mathcal{H}|\langle \delta'\phi,\hat{\Omega}\delta\phi\rangle=0, \forall  \delta\phi\in \mathcal{S}\}.
\end{equation}
Then it is not hard to show $\mathcal{S}^{\perp_\text{s}}=(\hat{\Omega}[\mathcal{S}])^\perp=\hat{\Omega}[\mathcal{S}^\perp]$, which further gives
\begin{equation}
    (\mathcal{S}^{\perp_\text{s}})^{\perp_\text{s}}=(\hat{\Omega}[\mathcal{S}^\perp])^{\perp_\text{s}}=\hat{\Omega}[(\hat{\Omega}[\mathcal{S}^\perp])^\perp]=\hat{\Omega}[\hat{\Omega}[(\mathcal{S}^\perp)^\perp]=(\mathcal{S}^\perp)^\perp=\bar{\mathcal{S}},
\end{equation}
where the bar represents the closure in $\mathcal{H}$. Since any subspace is dense in its closure,  below we shall not bother ourselves by saying that the double symplectic complement of any subspace is itself.

\subsection{Canonical energy, explicit expression at null infinity, and relation to second order perturbations}\label{subcanonical}
Finally, associated with an arbitrary background symmetry generated by  $\kappa^a$, we would like to introduce the corresponding canonical energy, which is a bilinear form on the space of linear on-shell perturbations defined as
\begin{equation}
    \mathcal{E}_\kappa(\delta_1\phi,\delta_2\phi)=\Omega_{AB}\delta_1\phi^A\mathscr{L}_\kappa\delta_2\phi^B.
\end{equation}
It is easy to show that not only is the canonical energy conserved and symmetric, but also gauge invariant.

\begin{figure}
	\centering
	\includegraphics[width=15cm]{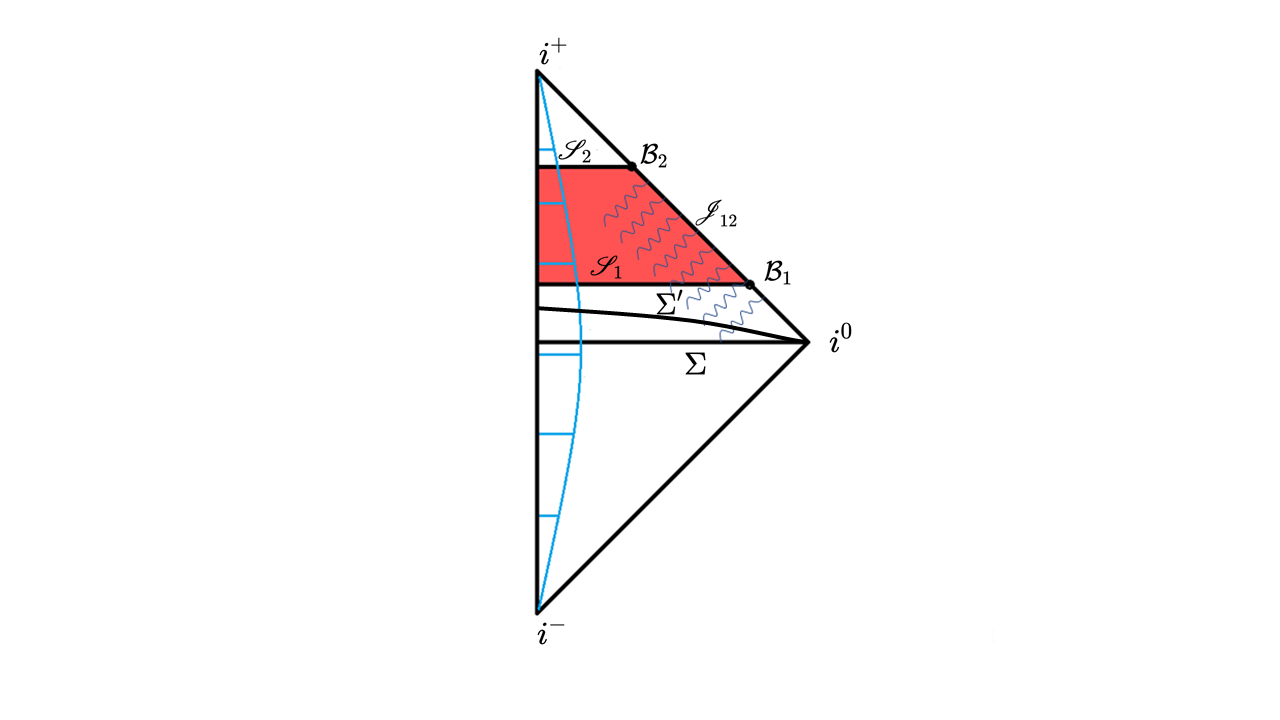}
	\caption{Penrose diagram for a perturbed relativistic star in an asymptotically flat spacetime.  $\Sigma$ and $\Sigma'$ are spacelike Cauchy surface terminating at the spatial infinity $i^0$ and the boundary of the red region is given by spacelike hypersurfaces $\mathscr{S}_1$ and $\mathscr{S}_2$ as well as the portion of the null infinity $\scri_{12}$, which is bounded by its two cross-sections $\mathcal{B}_1$ and $\mathcal{B}_2$.}
	\label{null}
\end{figure}

To obtain the expression of the canonical energy $\mathcal{E}_t(\delta_1\phi,\delta_2\phi)$ at the future null infinity $\scri$ for later use, we like to introduce the unphysical metric $\tilde{g}_{ab}=\Omega^2g_{ab}$ with the conformal factor $\Omega=0$ corresponding to the location of $\scri$ and work in the gauge such that $\tilde{\nabla}_an_b|_\scri=n^bA_b|_\scri=0$ with $n_b=\tilde{\nabla}_b\Omega$. Here the indices are raised or lowered by our unphysical metric, which is smooth across $\scri$. We further assume that $\Omega$ near $\scri$ and $\tilde{g}_{ab}$ at $\scri$ are universal quantities in the sense that $\delta\Omega=0$ near $\scri$ and $\delta\tilde{g}_{ab}|_\scri=0$. Whence the symplectic potential at $\scri$ for the electromagnetic part can be written as
\begin{equation}
    \bm{\theta}_{EM}=-F^{ab}\delta A_b\tilde{\epsilon}_{aefg}=\Pi^b\delta A_b\hat{\tilde{\bm{\epsilon}}}
\end{equation}
with $\Pi^b=n_aF^{ab}$ and the induced volume $3$-form given by $\tilde{\bm{\epsilon}}=-\bm{n}\wedge\hat{\tilde{\bm{\epsilon}}}$. Whence the symplectic current reads
\begin{equation}
    \bm{\omega}_{EM}=(\delta_2A_b\delta_1\Pi^b-\delta_1A_b\delta_2\Pi^b)\hat{\tilde{\bm{\epsilon}}},
\end{equation}
where one should keep in mind that $\delta \Pi^b=n_a\tilde{g}^{ac}\tilde{g}^{bd}\delta F_{cd}$.
Note that $t^a=cn^a$ at $\scri$ with $c$ a positive constant for the future directed timelike Killing vector field $t^a$, so we have $\mathscr{L}_t\delta \bm{A}=d(t\cdot\delta\bm{A})+t\cdot d\delta \bm{A}=\sigma \bm{n}+cn\cdot \delta \bm{F}$ at $\scri$ for some function $\sigma$. In addition, $0=\tilde{\nabla}_an^a\tilde{\bm{\epsilon}}=\mathscr{L}_n\tilde{\bm{\epsilon}}=-\bm{n}\wedge\mathscr{L}_n\hat{\tilde{\bm{\epsilon}}}$ at $\scri$ implies that $\mathscr{L}_n\hat{\tilde{\bm{\epsilon}}}=0$ when restricted onto $\scri$. As a result, we have
\begin{equation}\label{emnull}
\begin{aligned}
    \mathcal{E}^{EM}_t(\delta_1\phi,\delta_2\phi)&=\int_{\scri_{12}} c[2\delta_1\Pi_b\delta_2\Pi^b-\mathscr{L}_n(\delta_1 A_b\delta_2\Pi^b)]\hat{\tilde{\bm{\epsilon}}}\\
    &=2c\int_{\scri_{12}} \delta_1\Pi_b\delta_2\Pi^b\hat{\tilde{\bm{\epsilon}}}-c\int_{\scri_{12}}\mathscr{L}_n(\delta_1 A_b\delta_2\Pi^b\hat{\tilde{\bm{\epsilon}}})\\
    &=2c\int_{\scri_{12}} \delta_1\Pi_b\delta_2\Pi^b\hat{\tilde{\bm{\epsilon}}}+c(\int_{\mathcal{B}_2}\delta_1 A_b\delta_2\Pi^b\bm{\varepsilon}-\int_{\mathcal{B}_1}\delta_1 A_b\delta_2\Pi^b\bm{\varepsilon})\\
     &=2c\int_{\scri_{12}} \delta_1\Pi_b\delta_2\Pi^b\hat{\tilde{\bm{\epsilon}}}+c(\int_{\mathcal{B}_2}\delta_1 A^b\mathscr{L}_n\delta_2 A_b\bm{\varepsilon}-\int_{\mathcal{B}_1}\delta_1 A^b\mathscr{L}_n\delta_2 A_b\bm{\varepsilon})\\
    \end{aligned}
\end{equation}
where the portion of the null infinity $\scri_{12}$, as depicted in Fig. \ref{null}, is bounded by its two cross sections $\mathcal{B}_1$ and $\mathcal{B}_2$ with $\bm{\varepsilon}\equiv-n\cdot\hat{\tilde{\bm{\epsilon}}}$ the induced volume $2$-form on them. Similarly, the canonical energy for the gravitational part can be obtained as\cite{HW}
\begin{equation}\label{grnull}
\begin{aligned}
    \mathcal{E}^g_t(\delta_1\phi,\delta_2\phi)&=c\int_{\scri_{12}} \delta_1 N_{ab}\delta_2 N^{ab}\hat{\tilde{\bm{\epsilon}}}-\frac{c}{2}(\int_{\mathcal{B}_2} \tau_1^{ab}\delta_2 N_{ab}\bm{\varepsilon}-\int_{\mathcal{B}_1} \tau_1^{ab}\delta_2 N_{ab}\bm{\varepsilon})\\
    &=c\int_{\scri_{12}} \delta_1 N_{ab}\delta_2 N^{ab}\hat{\tilde{\bm{\epsilon}}}+\frac{c}{2}(\int_{\mathcal{B}_2} \tau_1^{ab}\mathscr{L}_n\tau_{2ab}\bm{\varepsilon}-\int_{\mathcal{B}_1} \tau_1^{ab}\mathscr{L}_n\tau_{2ab}\bm{\varepsilon}),
    \end{aligned}
\end{equation}
where $N_{ab}$ is the well known Bondi news tensor and its variation at $\scri$ satisfies
\begin{equation}
    \delta N_{ab}=-\mathscr{L}_n\tau_{ab}+4\Omega^{-1}n_{(a}\tau_{b)c}n^c-\Omega^{-1}n^cn_c\tau_{ab}
\end{equation}
with $\tau_{ab}\equiv\Omega\delta g_{ab}$. The canonical energy for the charged fluid part is automatically zero at null infinity since our charged star has a spatial compact support.

On the other hand, to develop the relation of the canonical energy with the second order perturbations, we would like to work in the gauge with the Lagrangian displacement vanishing to all orders. Accordingly, we have
\begin{equation}\label{relationtosecond}
\begin{aligned}
\mathcal{E}_\kappa(\delta\phi,\delta\phi)&=\int_\Sigma[\bm{\omega}^g(g_{ab};\Delta g_{ab},\mathscr{L}_\kappa\Delta g_{ab})+\bm{\omega}^{EM}(A_a;\Delta A_a,\mathscr{L}_\kappa \Delta A_a)]\\
&=\int_\Sigma [\Delta\bm{\omega}^g(g_{ab};\Delta g_{ab},\mathscr{L}_\kappa g_{ab})+\Delta \bm{\omega}^{EM}(A_a;\Delta A_a,\mathscr{L}_\kappa A_a)]\\
&=\Delta^2\mathcal{Q}_\kappa+\int_\Sigma \Delta \{\kappa^a \Delta[ (2G^b{}_a-T_{EM}^b{}_a-\nabla_c F^{cb}A_a)\epsilon_{befg}]+\kappa^a [(-G^{bc}+\frac{1}{2}T_{EM}^{bc})\Delta g_{bc}+\nabla_bF^{bc}\Delta A_c]\epsilon_{aefg}\}\\
&=\delta^2\mathcal{Q}_\kappa +\int_\Sigma\kappa^a\Delta \{[\Delta (T_m^b{}_a\epsilon_{befg})-\frac{1}{2}T_m^{bc}\Delta g_{bc}\epsilon_{aefg}]+[\Delta (J^bA_a\epsilon_{befg})-J^c\Delta A_c\epsilon_{aefg}]\}.
\end{aligned}
\end{equation}
Here we have used Eq. (\ref{fi}) in the third step by replacing $\bm{C}_a$ and $\bm{E}$ with the corresponding terms purely from gravitational and electromagnetic fields. In addition, we have also used the fact that the ADM charge is gauge invariant in the last step.

\section{Physically stationary perturbations and dynamical stability }\label{DS}
Dynamical stability we are concerned with is mode stability. That is to say, our charged star in dynamic equilibrium is mode stable if there does not exist any non-pure-gauge linearized solution with the time dependence of the form $e^{\Lambda t}$ with $\Lambda>0$. Otherwise, it is mode unstable. Compared to a complete analysis of linearized perturbation equations,  one favorable way of proving mode stability is to construct a positive-definite conserved norm on the space of linear on-shell perturbations, because this precludes those perturbations with exponential growth. A nice candidate for such a norm is the canonical energy $\mathcal{E}_t$. To be more specific, let us first introduce the physically stationary perturbations. A linear on-shell perturbation $\delta\phi$ off of a stationary background is called physically stationary if
\begin{equation}
\begin{aligned}
    \mathscr{L}_t(\delta g_{ab}+\mathscr{L}_X g_{ab})&=0,\\
    \mathscr{L}_t(\delta A_a+\mathscr{L}_X A_a+\nabla_a\vartheta)&=0,\\
    \mathscr{L}_t(\delta \bm{\mathcal{N}}+\mathscr{L}_X\bm{\mathcal{N}})=-\mathscr{L}_{[t,\xi-X]}\bm{\mathcal{N}}&=0\\
    \mathscr{L}_t(\delta s+\mathscr{L}_Xs)=-\mathscr{L}_{[t,\xi-X]}s&=0
    \end{aligned}
\end{equation}
with $X^a$ the asymptotic symmetry generator and $\vartheta$ satisfying $n^a\tilde{\nabla}_a\vartheta|_\scri=0$, or equivalently
\begin{equation}\label{stationary}
    \mathscr{L}_t\delta \phi=-\mathscr{L}_{[t,X]}\phi-\delta_{\mathscr{L}_t\vartheta}\phi+\text{trivial}.
\end{equation}
For the gravitational part, we first note that the news tensor vanishes for all stationary spacetimes\cite{Geroch,WZ}, so the perturbed news tensor induced by $\delta g_{ab}+\mathscr{L}_X g_{ab}$ vanishes. In addition, the news tensor keeps vanishing under the asymptotic symmetry transformation on a stationary spacetime, so the perturbed news tensor induced by $\mathscr{L}_X g_{ab}$ also vanishes. As a result, the perturbed news tensor vanishes for a physically stationary perturbation. While for the electromagnetic part, we first note that $\mathscr{L}_t \bm{A}=0$ implies $n\cdot\bm{F}\propto \bm{n}$ at $\scri$. Moreover, the commutator $[t,X]^a$, at most, leads to an asymptotic spatial translation, which is proportional to $n^a$ at the null infinity. So we have $\mathscr{L}_{[t,X]}\bm{A}\propto \bm{n}$ at $\scri$. On the other hand, $\mathscr{L}_t\vartheta|_\scri=cn^a\tilde{\nabla}_a\vartheta=0$, so we have $\tilde{\nabla}_a\mathscr{L}_t\vartheta\propto n_a$ at $\scri$. Then by Eq. (\ref{stationary}), for a physically stationary perturbation, we have $\mathscr{L}_n \delta A_a|_\scri\propto n_a$, which vanishes when restricted onto $\scri$. Accordingly, according to Eq. (\ref{emnull}) and Eq. (\ref{grnull}), we have that the canonical energy $\mathcal{E}_t(\delta\phi,\delta\phi)$ across the portion of the null infinity is apparently non-negative for both the gravitational part and the electromagnetic one if $\mathcal{B}_2$ is where a physically stationary perturbation solution settles down and $\mathcal{B}_1$ is chosen at the spatial infinity where the boundary term is supposed to vanish.
In addition, the canonical energy $\mathcal{E}_t(\delta\phi,\delta_\text{ps}\phi)$ across the null infinity vanishes for any physically stationary perturbation $\delta_\text{ps}\phi$, which implies
\begin{equation}\label{misspoint}
   \mathcal{E}_t(\delta\phi,\delta_\text{ps}\phi)= \mathcal{E}_t(\mathscr{S}_1;\delta\phi,\delta_\text{ps}\phi)=\mathcal{E}_t(\mathscr{S}_2;\delta\phi,\delta_\text{ps}\phi)
\end{equation}
with the canonical energy evaluated on $\mathscr{S}_i$ denoted by $\mathcal{E}_t(\mathscr{S}_i;\delta\phi,\delta_\text{ps}\phi)$.

Next let us define the subspace $\mathcal{V}\subset\mathcal{H}$ of the linear on-shell perturbations such that $\delta\phi\in \mathcal{V}$ satisfies
\begin{equation}
    \mathcal{E}_t(\delta\phi,\delta_{\text{ps}}\phi)=0
\end{equation}
for any physically stationary perturbation $\delta_{\text{ps}}\phi$. This amounts to saying that $\mathcal{V}$ is the symplectic complement of the subspace $\mathcal{W}$ spanned by the perturbations of the form $\mathscr{L}_t\delta_\text{ps}\phi$, i.e.,
\begin{equation}\label{conditions}
  \delta H_{\mathscr{L}_t\vartheta}=\langle \delta \phi,\hat{\Omega}\delta_{\mathscr{L}_t\vartheta}\phi\rangle=0,\quad   \delta H_{[t,X]}=\langle \delta\phi, \hat{\Omega}\mathscr{L}_{[t,X]}\phi\rangle=0, \quad  \langle \delta\phi,\hat{\Omega}[\text{trivial}]\rangle=0,
\end{equation}
where the first requirement is trivial since it is automatically satisfied, and the second one requires the perturbation should not change the ADM $3$-momentum. Note that the double symplectic complement of $\mathcal{W}$ is itself, thus the symplectic completement of $\mathcal{V}$ in itself is $\mathcal{V}\cap \mathcal{W}$, which implies that the degeneracy of the canonical energy is given precisely by the physically stationary perturbations when restricted onto $\mathcal{V}$. With this observation, we have the following criterion for the dynamical stability of our charged star.

\textbf{Criterion for dynamical stability}: If $\mathcal{E}_t(\delta\phi,\delta\phi)\ge 0$ for each  $\delta\phi\in \mathcal{V}$, then our charged star has mode stability with respect to the perturbations within $\mathcal{V}$. On the other hand, if there exists some $\delta\phi\in \mathcal{V}$ such that $\mathcal{E}_t(\delta\phi,\delta\phi)<0$, then our charged star has instability in the sense that such a perturbation cannot approach a physically stationary perturbation at late times.

It is noteworthy that  the physically stationary perturbations play a vital role in the above criterion. For the stability criterion, $\mathcal{E}_t(\delta\phi,\delta\phi)=0$ only for the physically stationary perturbations in $\mathcal{V}$. This can be argued by contradiction. Suppose that we have $\mathcal{E}_t(\delta\phi,\delta\phi)=0 $ for a perturbation $\delta\phi\in\mathcal{V}$, which is not a physically stationary one. Then there exists another perturbation $\delta'\phi\in\mathcal{V}$ such that $\mathcal{E}_t(\delta'\phi,\delta\phi)\neq 0$. Whence we have
\begin{equation}
    \mathcal{E}_t(\delta'\phi+z\delta\phi,\delta'\phi+z\delta\phi)=\mathcal{E}_t(\delta'\phi,\delta'\phi)+2z\mathcal{E}_t(\delta'\phi,\delta\phi),
\end{equation}
which can become negative by an appropriate choice of $z$. But this contradicts with the assumption that the canonical energy $\mathcal{E}_t$ is non-negative in $\mathcal{V}$. Thus for those perturbations which are not physically stationary ones, the canonical energy $\mathcal{E}_t$ provides a positive definite conserved norm during the evolution from one Cauchy surface $\Sigma$ to a later one $\Sigma'$, as depicted in Fig. \ref{null},  guaranteeing the mode stability. On the other hand, a physically stationary perturbation is stable by definition, albeit its zero canonical energy.
 While for the instability criterion, we can also argue for it by contradiction. Suppose that the negative canonical energy mode $\delta\phi$ approaches a physically stationary solution $\delta_\text{ps}\phi$ at late times, say $\mathscr{S}_2$ in Fig. \ref{null}, then we have
\begin{equation}
    \mathcal{E}_t(\mathscr{S}_2;\delta_\text{ps}\phi,\delta_\text{ps}\phi)\le \mathcal{E}_t(\delta\phi,\delta\phi)<0
\end{equation}
 due to the non-negative net flux across the null infinity. On the other hand,  Eq. (\ref{misspoint}) gives rise to
 \begin{equation}
\mathcal{E}_t(\mathscr{S}_2;\delta_\text{ps}\phi,\delta_\text{ps}\phi)=\mathcal{E}_t(\delta\phi,\delta_\text{ps}\phi)=0
 \end{equation}
 according to the definition of $\mathcal{V}$.  This leads to an obvious contradiction. So we are done.

However, such a criterion has its apparent shortcoming because one can only talk about the stability of the perturbations within the subspace $\mathcal{V}$.
Gratefully, the second condition in Eq. (\ref{conditions}) imposed on $\mathcal{V}$ can always be achieved by adding a perturbation generated by the Lorentz boosts on the background. Since such an added perturbation is stable, the above criterion for the stability of the perturbations in $\mathcal{V}$ also implies the criterion for the stability of the perturbations in the larger subspace with the second condition in Eq. (\ref{conditions}) disregarded. On the other hand, regarding the third condition in Eq. (\ref{conditions}),  we also have good news for both non-axisymmetric and axisymmetric perturbations, on which we shall elaborate separately below.

\subsection{Nonaxisymmetric perturbations}
As to the neutral star, Friedman shows that the aforementioned third condition does not lead to a real physical restriction for nonaxisymmetric perturbations, because it can be achieved in the suitable background solution by adding a trivial perturbation\cite{F1978}. Here we focus only onto the background in which $\nabla_as\neq0$ to show the key strategy developed in \cite{F1978} for the neutral star as well as its utterly equal applicability to our charged star.

First, by working in the gauge $\delta\phi=(\Delta g_{ab}, \Delta A_a,0)$, we have
\begin{equation}
\begin{aligned}
    \langle \delta\phi,\hat{\Omega}(0,0,\tilde{\eta})&=\int_\Sigma\tilde{\eta}^a\Delta P_{aefg}\\
    &=\int_\Sigma \frac{1}{n^2}\mathcal{N}^{abc}\nabla_bZ_c\Delta P_{aefg}\\
    &=\int_\Sigma\nabla_bZ_c\Delta(\frac{1}{n^2}\mathcal{N}^{abc}P_{aefg})\\
    &=6\int_\Sigma \nabla_{[e}Z_f\Delta(\frac{\rho+p}{n}u_{g]}+A_{g]})\\
    &=\int_\Sigma d\bm{Z}\wedge\Delta (\frac{\rho+p}{n}\bm{u}+\bm{A})\\
    &=\int_\Sigma \bm{Z}\wedge \Delta d(\frac{\rho+p}{n}\bm{u}+\bm{A})\\
    &=\int_\Sigma F\Delta [ds\wedge d(\frac{\rho+p}{n}\bm{u}+\bm{A})],
    \end{aligned}
\end{equation}
where we have used $\Delta(\frac{1}{n^2}\mathcal{N}^{abc})= u^{[a}\Lambda^{bc]}$ for some antisymmetric tensor $\Lambda^{bc}$ and $u\cdot d\bm{Z}=u^a\bm{P}_a=0$ in the third step. We define the vorticity and circulation in the presence of the electromagnetic field as $\bm{\omega}=d(\frac{\rho+p}{n}\bm{u}+\bm{A})$ and $\bm{\Gamma}=ds\wedge\bm{\omega}$, respectively. Then we have
\begin{equation}
    \begin{aligned}
        u\cdot\bm{\omega}&=u\cdot[d(\frac{\rho+p}{n})\wedge \bm{u}+\frac{\rho+p}{n}d\bm{u}+\bm{F}]\\
        &=d(\frac{\rho+p}{n})+\frac{\rho+p}{n}u^a\nabla_au_b+u^aF_{ab}\\
        &=d(\frac{\rho+p}{n})-\frac{dp}{n}=Tds,
    \end{aligned}
\end{equation}
where we have used the first equation of Eq. (\ref{conservationlaw}) as well as the fact that both the particle number density and the entropy per particle along the divergence free $4$-velocity are constant for our charged star in dynamic equilibrium. Whence we obtain Ertel's theorem
\begin{equation}
    \mathscr{L}_u\bm{\omega}=dT\wedge ds
\end{equation}
  as well as $u\cdot\bm{\Gamma}= \mathscr{L}_u\bm{\Gamma}=0$ in the presence of the electromagnetic field, which implies $\bm{\Gamma}=\gamma \bm{\mathcal{N}}$ and $\mathscr{L}_u\gamma=0$. On the other hand, note that
 \begin{equation}
     \langle (0,0,\tilde{\eta}'),\hat{\Omega}(0,0,\tilde{\eta})\rangle=\int_\Sigma F\mathscr{L}_{\tilde{\eta}'}\bm{\Gamma}
 \end{equation}
 for any trivial displacement $\tilde{\eta}'^a=\frac{1}{2n^2}\mathcal{N}^{abc}(dF'\wedge ds)_{bc}$. So we are required to ask whether there exists a trivial displacement such that $\mathscr{L}_{\tilde{\eta}'}\bm{\Gamma}=-\Delta \bm{\Gamma}$ on $\Sigma$, i.e.,
 \begin{equation}
 \begin{aligned}
     -\Delta\gamma&=\mathscr{L}_{\tilde{\eta}'}\gamma=\frac{1}{n^2}\mathcal{N}^{abc}(dF')_b (ds)_c (d\gamma)_a=\frac{1}{n}\epsilon^{abcd}(d\gamma)_a(ds)_b(dF')_cu_d\\
     &=\frac{\frac{\partial F'}{\partial\varphi}}{n}\epsilon^{abcd}(d\gamma)_a(ds)_b[(d\varphi)_c-\Omega(dt)_c]u_d=\frac{\frac{\partial F'}{\partial\varphi}}{n}\epsilon^{abcd}(d\gamma)_a(ds)_b(d\varphi)_c(dt)_d(u_t+\Omega u_\varphi)\\
     &=\frac{\frac{\partial F'}{\partial\varphi}}{n}\epsilon^{abcd}(d\gamma)_a(ds)_b(d\varphi)_c(dt)_d(g_{tt}+\Omega g_{t\varphi}+\Omega g_{\varphi t}+\Omega^2g_{\varphi\varphi})/|v|=-\frac{|v|\frac{\partial F'}{\partial\varphi}}{n}\epsilon^{abcd}(d\gamma)_a(ds)_b(d\varphi)_c(dt)_d,
     \end{aligned}
 \end{equation}
where we have assumed that  $(\gamma, s, t, \varphi)$ can offer a local atlas for our charged star with $t^a(dt)_a=\varphi^a(d\varphi)_a=1$ and used $\mathscr{L}_uF'=0$. With this,  the answer is definitely yes, because one can always find a $F'$ on $\Sigma$ such that
\begin{equation}
    \frac{\partial F'}{\partial\varphi}=\frac{n\Delta \gamma}{|v|\epsilon^{abcd}(d\gamma)_a(ds)_b(d\varphi)_c(dt)_d}.
\end{equation}
 The upshot is that the third condition in Eq. (\ref{conditions}) does not correspond to a real physical restriction on the nonaxisymmetric perturbations in most of the cases we are interested in, since it can be fulfilled by supplementing a trivial perturbation.

\subsection{Axisymmetric perturbations}
For an arbitrary axisymmetric perturbation $\delta \mathcal{Q}=(\delta g_{ab},\delta A_a,\delta\bm{\mathcal{N}},\delta s)$, which may not be described by our Lagrangian formulation, one can follow exactly the same construction developed in \cite{GSW} for the neutral star to show\footnote{The derivation presented in \cite{GSW} has used $u^a\nabla_a s=0$ and $u^a\nabla_aj=0$ for axisymmetric perturbations. The validity of the former is obvious since it holds actually for arbitrary perturbations. The validity of the latter also for our charged star can be seen through $0=\varphi_b(\nabla_aT_m^{ab}-F^{bc}J_c)=\nabla_a(jJ^a+p\varphi^a)=\nabla_a(jJ^a)=J^a\nabla_aj$.}
\begin{equation}
    \mathscr{L}_t\delta\bm{\mathcal{N}}=-\mathscr{L}_\xi\bm{\mathcal{N}},\quad \mathscr{L}_t\delta s=-\mathscr{L}_\xi s,\quad \mathscr{L}_t\delta j=-\mathscr{L}_\xi j,
\end{equation}
where  $\xi^a=|v|\delta u^a+\beta \varphi^a$ with an arbitrary axisymmetric scalar $\beta$ satisfying $u^a\nabla_a\beta=\delta u^a\nabla_a\Omega$. This means that $\mathscr{L}_t\delta\mathcal{Q}$ can be realized in our Lagrangian description as $\hat{\delta}\phi=(\mathscr{L}_t \delta g_{ab},\mathscr{L}_t\delta A_a,\xi^a)$ with $\Delta j=0$. Now for an arbitrary axisymmetric trivial displacement $\tilde{\eta}^a$, we have
\begin{equation}\label{v1}
\begin{aligned}
    \langle \mathscr{L}_t\hat{\delta}\phi,\hat{\Omega}(0,0,\tilde{\eta})\rangle&= -\langle \hat{\delta}\phi,\hat{\Omega}(0,0,\mathscr{L}_t\tilde{\eta})\rangle
    =-\langle \hat{\delta}\phi,\hat{\Omega}(0,0,\gamma\varphi)\rangle
    	=\int_\Sigma \gamma\varphi^a\Delta P_{abcd}\\
		                          &=\int_\Sigma \gamma\varphi^a\Delta[\frac{\rho+p}{n}(n\epsilon_{abcd}+u_a\mathcal{N}_{bcd})-A_fJ^f\epsilon_{abcd}+A_a\mathcal{N}_{bcd}]\\
		                          &=\int_\Sigma \gamma \mathcal{N}_{bcd}\Delta[(\frac{\rho+p}{n}u_a+A_a)\varphi^a]=\int_\Sigma \gamma \bm{\mathcal{N}}\Delta j=0,
	\end{aligned}
\end{equation}
where we have used $\mathscr{L}_t\tilde{\eta}=\gamma \varphi^a$ for some function $\gamma$\cite{GSW}. Similarly, one can also have
\begin{equation}\label{v2}
    \langle \mathscr{L}_t\hat{\delta}\phi,\hat{\Omega}\mathscr{L}_{[t,X]}\phi\rangle=-\langle \hat{\delta}\phi,\hat{\Omega}\mathscr{L}_Y\phi\rangle=0,
\end{equation}
where we have used the fact that $Y^a=[t,[t,X]]^a$ vanishes at infinity for any asymptotic symmetry generator $X^a$.

Eq. (\ref{v1}) and Eq. (\ref{v2}) imply that $\mathscr{L}_t^2\delta\mathcal{Q}\in \mathcal{V}$ for an arbitrary axisymmetric perturbation $\delta\mathcal{Q}$. On the other hand, the mode stability for $\delta\mathcal{Q}$ is equivalent to the mode stability for $\mathcal{L}_t^2\delta\mathcal{Q}$. Thus the mode stability for perturbations within $\mathcal{V}$ implies the mode stability for all perturbations in axisymmetric case.

\section{Thermodynamic stability and axisymmetric perturbations}\label{TS}
Our charged star in thermodynamic equilibrium is said to be linearly thermodynamically stable if $\delta^2S$ is negative for all the linear on-shell perturbations with $\delta\mathcal{M}=\delta N=\delta \mathcal{J}=\delta^2\mathcal{M}=\delta^2N=\delta^2\mathcal{J}=0$. Note that the variation of the first law of thermodynamics gives
\begin{equation}
    \delta^2 \mathcal{M}-\tilde{T}\delta^2S-\tilde{\mu}\delta^2N-\Omega\delta^2\mathcal{J}=\int_\Sigma (\delta \tilde{T}\delta(s\bm{\mathcal{N}})+\delta \tilde{\mu}\delta \bm{
\mathcal{N}}-\delta\Omega d\delta\bm{Q}_\varphi),
\end{equation}
which implies that the left side, denoted by $\mathcal{C}$ later on, depends solely on the first order variation of our system. Thus the criterion for the thermodynamic stability of our charged star is the positivity of $\mathcal{C}$ for all the linear on-shell perturbations with $\delta\mathcal{M}=\delta N=\delta\mathcal{J}=0$. In our Lagrangian description, we have $\delta N=\delta S=\delta^2 N=\delta^2 S=0$ automatically\footnote{We also have $\delta\mathcal{M}=0$ since we are always working within the subspace $\mathcal{H}$.}. Then by Eq. (\ref{relationtosecond}), we obtain
\begin{equation}
\mathcal{C}=\mathcal{E}_v-\int_\Sigma v^a\Delta \{[\Delta (T_m^b{}_a\epsilon_{befg})-\frac{1}{2}T_m^{bc}\Delta g_{bc}\epsilon_{aefg}]+[\Delta (J^bA_a\epsilon_{befg})-J^c\Delta A_c\epsilon_{aefg}]\}
\end{equation}
for the Killing field $v^a$, where $\mathcal{E}_v$ can be understood as the canonical energy in the comoving frame. We borrow directly from \cite{our} the following two identities
\begin{equation}
	\begin{aligned}
		u^a[J^c\Delta A_c \epsilon_{aefg}-\Delta(J^bA_a\epsilon_{befg})]&=-u\cdot\bm{ A} \Delta \bm{\mathcal{N}},\\
		u^a[\frac{1}{2}T_m^{bc}\Delta g_{bc}\epsilon_{aefg}-\Delta(T_m^b{}_a\epsilon_{befg})]&=\mu \Delta \bm{\mathcal{N}}+T\Delta (s\bm{\mathcal{N}}). \\
	\end{aligned}
\end{equation}
By the Lagrangian variation of these two identities, we further have
\begin{equation}
	\begin{aligned}
		u^a\Delta [J^c\Delta A_c \epsilon_{aefg}-\Delta(J^bA_a\epsilon_{befg})]&=0,\\
		u^a\Delta [\frac{1}{2}T^{bc}\Delta g_{bc}\epsilon_{aefg}-\Delta(T^b{}_a\epsilon_{befg})]&=0,\\
	\end{aligned}
\end{equation}
where we have used $\Delta u^a\propto u^a$ as well as the fact that any order of the Lagrangian variation of $\bm{\mathcal{N}}$ and $s\bm{\mathcal{N}}$ vanishes. Accordingly, we obtain
\begin{equation}
    \mathcal{C}=\mathcal{E}_v,
\end{equation}
whereby we further have
\begin{equation}
    \mathcal{C}=\mathcal{E}_t
\end{equation}
for axisymmetric perturbations.  With this observation, we obtain the following criterion for the thermodynamic stability of our charged star.

\textbf{Criterion for thermodynamic stability}: For our charged star in thermodynamic equilibrium, the necessary condition for its thermodynamic stability is the positivity of $\mathcal{E}_v$ for all the linear on-shell perturbations with $\delta\mathcal{J}=0$, and the necessary condition for its thermodynamic stability with respect to all the linear on-shell axisymmetric perturbations with $\delta\mathcal{J}=0$ is the positivity of the corresponding $\mathcal{E}_t$.

Here the necessary condition can be replaced by the necessary and sufficient condition if and only if $\frac{\delta s}{|D_as|}$ is bounded for all allowable perturbations of $s$, since only in this case can the perturbations be implemented within our Lagrangian description\cite{F1978}. One such an application scenario is provided by the so-called isentropic star, which is regarded generally as a good approximation at least for the convective part of a real life star\cite{Weinberg}. For such an isentropic star, we have a uniform entropy per particle throughout the whole star. As a result, $\frac{\delta s}{|D_as|}$ is bounded in the sense that not only do we have $D_as=0$ but also the allowable $\delta s=0$. In particular, let us consider the linear on-shell spherically symmetric perturbations of a static, spherically symmetric isentropic charged star, for which we have $\delta \mathcal{J}=0$ automatically. On the other hand, we also have no spherically symmetric trivial displacement because of
\begin{equation}
    0=\mathscr{L}_{\xi\frac{\partial}{\partial r}}\bm{\mathcal{N}}=d(\xi\frac{\partial}{\partial r}\cdot\bm{\mathcal{N}})=d(n\sqrt{h}\xi d\theta\wedge d\varphi)=\frac{\partial(n\sqrt{h}\xi)}{\partial r} dr\wedge d\theta\wedge d\varphi
\end{equation}
subject to the boundary condition $\xi=0$ at $r=0$ in the spherical coordinates. Together with the ADM $3$-momentum unchanged under the spherically symmetric perturbations, we know that there is no restriction on $\mathcal{V}$ for the spherically symmetric perturbations. Thus we end up with the corollary that the dynamic stability is equivalent to the thermodynamic stability for the spherically symmetric perturbations of the static, spherically symmetric isentropic charged star, which has also been obtained most recently in \cite{YFJ2021} instead by performing explicit calculations in the spherical coordinates.

\section{Discussions}
As promised, we have accomplished a thorough analysis of the dynamic and thermodynamic stability for the charged star. As we show explicitly in our paper, neither the presence of the electromagnetic field nor the Lorentz force experienced by the charged fluid makes any obstruction to the key steps towards the results obtained previously in \cite{GSW} for the neutral star. Thus our main results for the charged star are in close parallel with those presented in \cite{GSW} for the neutral star, but with various improvements scattered in our paper.

We conclude our paper with two interesting issues worthy of further investigation. The first one is related to the Gubser-Mitra conjecture. As alluded to in the introduction section, Hollands and Wald proved the Gubser-Mitra conjecture for black branes in AdS spacetimes by resorting to the criterion they established for the dynamic and thermodynamic stability of the black holes\cite{HW}. Thus it is tempting to ask whether one can obtain the similar equivalence between the dynamic instability and the thermodynamic instability for the AdS planar star by the same token. On the other hand, new progress has recently been made towards the Lagrangian formulation of dissipative fluids and its various implications\cite{CGL,GCL,LG,GLR,GGL}, so it is also interesting to explore what new insights one can gain by applying the Wald formalism to it. We hope to report some progress along these two lanes in the future.

\section*{Acknowledgements}
We are grateful to Bob Wald for his helpful communications regarding his work. We also thank Hu Zhu for his inspiring discussions on some issues related to this work.
This work is partly supported by the National Key Research and Development Program of China Grant No. 2021YFC2203001 as well as the NSFC under Grant Nos.
11975239, 12005088, 12035016, 12075026, and 12275350. JZ is also supported by the Beijing Research Fund for Talented Undergraduates.

\end{document}